\documentclass[twocolumn,english,aps,prx,showpacs,amsmath,amssymb,superscriptaddress]{revtex4}
\usepackage[latin9]{inputenc}
\usepackage{color}
\usepackage{amstext}
\usepackage{amssymb}
\usepackage{graphicx}

\makeatletter
\@ifundefined{textcolor}{}
{%
 \definecolor{BLACK}{gray}{0}
 \definecolor{WHITE}{gray}{1}
 \definecolor{RED}{rgb}{1,0,0}
 \definecolor{GREEN}{rgb}{0,1,0}
 \definecolor{BLUE}{rgb}{0,0,1}
 \definecolor{CYAN}{cmyk}{1,0,0,0}
 \definecolor{MAGENTA}{cmyk}{0,1,0,0}
 \definecolor{YELLOW}{cmyk}{0,0,1,0}
}

\usepackage{color}

\newcommand{\beq}{\begin{eqnarray}}
\newcommand{\eeq}{\end{eqnarray}}
\newcommand{\ys}[1]{\textcolor{black}{#1}}
\makeatother
\usepackage{babel}

\begin{document}

\title{Spin dynamics in NaFeAs and NaFe$_{0.53}$Cu$_{0.47}$As probed by resonant inelastic X-ray scattering}

\author{Yu Song}\thanks{Current affiliation: Department of Physics, Zhejiang University, Hangzhou 310027, China}

\email{yusong_phys@zju.edu.cn}

\selectlanguage{english}%

\affiliation{Department of Physics and Astronomy, Rice University, Houston, Texas 77005, USA}
\affiliation{Department of Physics, University of California, Berkeley, California 94720, USA}

\author{Weiyi Wang}

\affiliation{Department of Physics and Astronomy, Rice University, Houston, Texas 77005, USA}

\author{Eugenio Paris}
\affiliation{Photon Science Division, Swiss Light Source, Paul Scherrer Institut, 5232 Villigen PSI, Switzerland}

\author{Xingye Lu}
\affiliation{Center for Advanced Quantum Studies and Department of Physics, Beijing
	Normal University, Beijing 100875, China}
\affiliation{Photon Science Division, Swiss Light Source, Paul Scherrer Institut, 5232 Villigen PSI, Switzerland}

\author{Jonathan Pelliciari}\thanks{Current affiliation: National Synchrotron Light Source-II, Brookhaven National Laboratory, Upton, NY 11973, USA}
\affiliation{Photon Science Division, Swiss Light Source, Paul Scherrer Institut, 5232 Villigen PSI, Switzerland}

\author{Yi Tseng} 
\affiliation{Photon Science Division, Swiss Light Source, Paul Scherrer Institut, 5232 Villigen PSI, Switzerland}

\author{Yaobo Huang}
\affiliation{Photon Science Division, Swiss Light Source, Paul Scherrer Institut, 5232 Villigen PSI, Switzerland}

\author{Daniel McNally}
\affiliation{Photon Science Division, Swiss Light Source, Paul Scherrer Institut, 5232 Villigen PSI, Switzerland}

\author{Marcus Dantz}
\affiliation{Photon Science Division, Swiss Light Source, Paul Scherrer Institut, 5232 Villigen PSI, Switzerland}

\author{Chongde Cao}
\affiliation{MOE Key Laboratory of Materials Physics and Chemistry under Extraordinary Conditions and Shaanxi Key Laboratory of Condensed Matter Structures and Properties, School of Physical Science and Technology, Northwestern Polytechnical University, Xian 710072, China}

\author{Rong Yu}
\affiliation{Department of Physics and Beijing Key Laboratory of Opto-electronic Functional Materials and Micro-nano Devices, Renmin University of China, Beijing 100872, China}

\author{Robert J. Birgeneau}
\affiliation{Department of Physics, University of California, Berkeley, California 94720, USA}

\author{Thorsten Schmitt}
\affiliation{Photon Science Division, Swiss Light Source, Paul Scherrer Institut, 5232 Villigen PSI, Switzerland}

\author{Pengcheng Dai}

\email{pdai@rice.edu}

\selectlanguage{english}%

\affiliation{Department of Physics and Astronomy, Rice University, Houston, Texas
77005, USA}

\begin{abstract}

The parent compounds of iron-based superconductors are magnetically-ordered bad metals, with superconductivity appearing near a putative magnetic quantum critical point. The presence of both Hubbard repulsion and Hund's coupling leads to rich physics in these multiorbital systems, and motivated descriptions of magnetism in terms of itinerant electrons or localized spins.
The NaFe$_{1-x}$Cu$_x$As series consists of magnetically-ordered bad metal ($x=0$), superconducting ($x\approx0.02$) and magnetically-ordered semiconducing/insulating ($x\approx0.5$) phases, providing a platform to investigate the connection between superconductivity, magnetism and electronic correlations.
Here we use X-ray absorption spectroscopy and resonant inelastic X-ray scattering to study the valence state of Fe and spin dynamics in two NaFe$_{1-x}$Cu$_x$As compounds ($x=0$ and 0.47). We find that magnetism in both compounds arises from Fe$^{2+}$ atoms, and exhibits underdamped dispersive spin waves in their respective ordered states. 
The dispersion of spin excitations in NaFe$_{0.53}$Cu$_{0.47}$As is consistent with being quasi-one-dimensional. Compared to NaFeAs, the band top of spin waves in NaFe$_{0.53}$Cu$_{0.47}$As is slightly softened with significantly more spectral weight of the spin excitations. Our results indicate the spin dynamics in NaFe$_{0.53}$Cu$_{0.47}$As arise from localized magnetic moments
and suggest the iron-based superconductors are proximate to a correlated insulating state with localized iron moments.

\end{abstract}

\pacs{74.25.Ha, 74.70.-b, 78.70.Nx}

\maketitle

\section{Introduction}

The parent compounds of cuprate superconductors are magnetically-ordered Mott-insulators, where the spin-1/2 Cu$^{2+}$ moments forming antiferromagnetic (AF) order are completely localized and can be well-described by a Heisenberg Hamiltonian \cite{PLee2006,Scalapino,Coldea}. 
Because high-temperature superconductivity in cuprates is derived from 
 magnetically-ordered parent compounds after eliminating static AF order, 
it is generally believed that electronic correlations and magnetism are important
for the superconductivity of these materials  \cite{PLee2006,Scalapino}. Spin dynamics in the parent compounds of cuprate superconductors are described by spin waves due to interactions between localized moments \cite{Coldea}. Cuprate superconductors exhibit similar spin dynamics, especially in the underdoped regime and at high-energies, suggesting the inheritance of short-range spin correlations from their parent compounds \cite{Birgeneau2006,Sidis2007,Fujita2012,MPMDean2015}. In contrast, the parent compounds of iron-based superconductors (FeSCs) are magnetically-ordered bad metals with significant electronic correlations \cite{MQazilbash2009,ZYin2011,PDai2012,QSi2016}, suggesting that their spin dynamics may originate from either localized spins or itinerant electrons.

In contrast to the cuprates which can be effectively described using a single orbital, the FeSCs are multiorbital systems (with five $3d$ orbitals), and the presence of both Hubbard repulsion $U$ and Hund's coupling $J_{\rm H}$ leads to a rich theoretical phase diagram \cite{PDai2012,QSi2016}. At half-filling (with a total of $n=5$ electrons per Fe), both $U$ and $J_{\rm H}$ promote a Mott insulating state. Away from half-filling (such as $n=6$), while large $U$ and $J_{\rm H}$ lead to strong electronic correlations, a bad-metal state becomes favorable and a Mott insulating state is realized only for an intermediate range of $J_{\rm H}$ \cite{LdeMedici2011,LFanfarillo}. 

Since parent compounds of the FeSCs (with $n=6$) exhibit large values of resistivity, comparable to those of underdoped cuprates, they fall in the bad metal regime, and have been suggested to exhibit incipient Mott physics \cite{QSi2008}. In such a scenario, the metallic transport is due to the coherent part of the single-electronic excitations, while magnetism results from localized moments formed by the incoherent part \cite{QSi2016}. From this perspective, as electronic correlations are enhanced, a magnetically-ordered $n=6$ insulating phase should appear. BaFe$_2$Se$_3$ and BaFe$_2$S$_3$ tuned via pressure may be a realization of such a scenario \cite{HTakahashi2015,TYamauchi2015,JYing2017,SChi2016}, with their ground states transformed from magnetically-ordered insulators under ambient conditions to superconductivity under pressure (the strength of electronic correlations decreases with increasing pressure). Alternatively, the FeSCs may be anchored around a half-filled ($n=5$) Mott insulating state, with the magnetic and nematic orders in the parent compounds ($n=6$) being the competing states that suppress superconductivity \cite{LdeMedici2014}. 


\begin{figure}
	\includegraphics[scale=0.5]{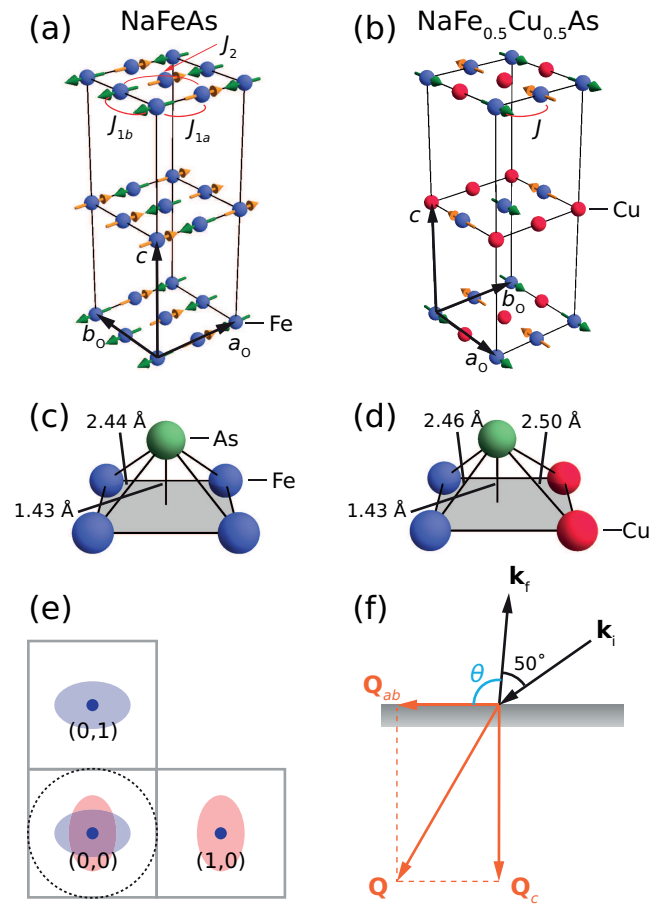} \protect\caption{(a) Magnetic structure of NaFeAs. (b) Magnetic structure of NaFe$_{0.5}$Cu$_{0.5}$As. (c) Local atomic environment in NaFeAs, with atomic distances from Ref. \cite{DParker2009} for $T=295$~K. (d) Local atomic environment in NaFe$_{0.5}$Cu$_{0.5}$As, with atomic distances from Ref. \cite{YSong2016} for $T=250$~K. (e) Reciprocal space of NaFeAs and NaFe$_{0.5}$Cu$_{0.5}$As. In both cases two types of domains are present, with spin waves exhibiting intensities at either ${\bf Q}=(1,0)$ or ${\bf Q}=(0,1)$. Since the reciprocal space probed by RIXS is limited to the dashed circle near ${\bf Q}=(0,0)$ for NaFeAs (2\% larger in diameter for NaFe$_{0.5}$Cu$_{0.5}$As), intensities from twin domains overlap. The shaded ellipsoids are schematic intensity contours for spin waves with anisotropic spin velocities in NaFeAs and NaFe$_{0.5}$Cu$_{0.5}$As. (f) Schematic scattering geometry of our RIXS measurements.}
	\label{Fig_schematic}
\end{figure}
 
As an archetypal parent compound of FeSCs, NaFeAs exhibits a tetragonal-to-orthorhombic transition below $T_{\rm S}\approx55$~K and stripe-type collinear magnetic order [Fig.~\ref{Fig_schematic}(a)] below $T_{\rm N}\approx45$~K \cite{DParker2009,SLi2009,JDWright2012}. The magnetic and structural phase transitions in NaFeAs are quickly suppressed upon Cu-doping, giving way to superconductivity near $x\approx0.02$ \cite{AWang2013}. With further increase of Cu concentration, NaFe$_{1-x}$Cu$_x$As starts to exhibit semiconducting/insulating electrical transport, accompanied by the appearance of short-range magnetic order and charge-gapped patches \cite{YSong2016,CYe2015}. Upon approaching $x=0.5$, Fe and Cu order into stripes and long-range magnetic order with $T_{\rm N}\approx200$~K reemerges [Fig.~\ref{Fig_schematic}(b)], accompanied by a charge gap that persists above $T_{\rm N}$ \cite{YSong2016,CMatt2016,ACharnuka2017,YZhou2020}. The ordering between Fe and Cu atoms in the $x=0.5$ compound is a structural analogue of the stripe-type magnetic order in NaFeAs, resulting in quasi-one-dimensional (quasi-1D) Fe chains separated by nonmagnetic Cu chains. Such a superstructure has been suggested to be essential for the \ys{insulating properties \cite{SZhang2017} and the electronic structure \cite{SLSkornyakov2021} of the $x=0.5$ compound.} The magnetic structure of NaFe$_{0.5}$Cu$_{0.5}$As can be obtained from that of NaFeAs simply by replacing half of Fe atoms by nonmagnetic Cu atoms and rotating the orientations of magnetic moments by 90$^\circ$, although with a much larger ordered moment \ys{($\geq 1.1$~$\mu_{\rm B}$/Fe \cite{YSong2016})} \ys{relative to that of NaFeAs ($\sim0.1$~$\mu_{\rm B}$/Fe \cite{DParker2009,SLi2009,JDWright2012})}. Locally, while the Fe-As distance is slightly larger in NaFe$_{0.5}$Cu$_{0.5}$As compared to NaFeAs, the heights of As atoms away from the transition metal planes are similar in the two compounds [Figs. 1(c) and (d)]. While simple valence counting suggests Fe to be in a $3+$ state with $n=5$ \cite{YSong2016},
the similar local environments of Fe ions in NaFeAs and NaFe$_{0.5}$Cu$_{0.5}$As suggests Fe in the latter to be in a $2+$ state with $n=6$.


Near $x\approx0.5$, the insulating/semiconducting electrical transport in NaFe$_{1-x}$Cu$_x$As exhibits a low-temperature resistivity $\sim 1$~$\Omega\cdot \rm{cm}$. Such a resistivity value is comparable to that found in Fe$_{1+\delta-x}$Cu$_x$Te (0.1-1~$\Omega\cdot{\rm cm}$ when $x\approx0.5$) \cite{AVaipolin1992,FAbdullaev2006,DZocco2012,PValdivia2015}, but contrasts with the much smaller resistivity values ($\sim 1$~${\rm m}\Omega\cdot{\rm cm}$) in the Sr(Fe$_{1-x}$Cu$_x$)$_2$As$_2$ \cite{YYan2013} and Ba(Fe$_{1-x}$Cu$_x$)$_2$As$_2$ \cite{WWang2017} series. Such a difference may result from stronger electronic correlations in NaFeAs and FeTe \cite{ZYin2011,MYi2017}, which persist upon $\sim 50\%$ Cu-doping and cause these systems to exhibit larger resistivity values. Another possibility is that while the structure of Sr(Fe$_{1-x}$Cu$_x$)$_2$As$_2$ and Ba(Fe$_{1-x}$Cu$_x$)$_2$As$_2$ allows the formation of $c$-axis collapsed phases with As-As covalent bonds \cite{AKressig2008,VKAnand2012}, such a phase is unlikely in NaFe$_{1-x}$Cu$_x$As and Fe$_{1+\delta-x}$Cu$_x$Te. From a theoretical perspective, insulating/semiconducting NaFe$_{1-x}$Cu$_x$As with $x\approx0.5$ may be a Mott insulator \cite{YSong2016}, in which an electronic gap opens due to strong electron-electron interactions; alternatively, a Slater insulator scenario, in which an electronic gap results from the symmetry lowering associated with magnetic order, has also been suggested \cite{ACharnuka2017}. In the latter scenario, persistent insulating/semiconducting above $T_{\rm N}$ is suggested to result from spin fluctuations that persist well above $T_{\rm N}$ in a quasi-1D system, which lead to an electronic gap similar to static magnetic order.

\begin{figure}
	\includegraphics[scale=0.62]{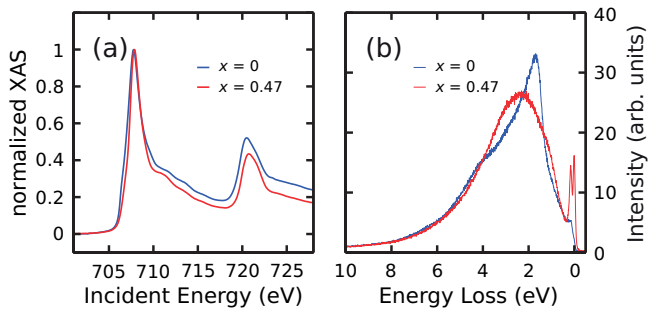} \protect\caption{(Color online) (a) X-ray absorption spectroscopy measurements for NaFe$_{1-x}$Cu$_x$As for $x=0$ and $x=0.47$. The measured spectra have been normalized so that the intensity maximum (minimum) corresponds to 1 (0). (b) Typical RIXS spectra compared for $x=0$ \ys{(${\bf Q}_{ab}=(0.50,0)$)} and $x=0.47$ \ys{(${\bf Q}_{ab}=(0.51,0)$)} samples. All RIXS spectra are normalized by integrated intensity within the energy loss range $[0.4,10]$~eV.}
	\label{Fig_XAS}
\end{figure} 

Spin dynamics in the FeSCs not only contain information on their magnetic interactions, but also provide clues regarding the size of the fluctuating magnetic moment (seen within timescale of the experimental probe), which in turn reflects the system's degree of itinerancy/localization \cite{PDai2015,NMannella,CWatzenbock}. 
While spin dynamics in FeSCs and related compounds are typically investigated using inelastic neutron scattering (INS) \cite{PDai2015,DSInosov2016,CZhang2014}, resonant inelastic X-ray scattering (RIXS) has emerged as a useful complementary probe \cite{LJPAment2011,KZhou2013,CMonney2013,HGretarsson2015,JPelliciari2016,TNomura2016,JPelliciari2016_2,JPelliciari2017,CRhan2019,FGarcia2019,JPelliciari2019}. 
Since NaFeAs and NaFe$_{0.5}$Cu$_{0.5}$As exhibit significant similarities in their crystal and magnetic structures, a comparison of the spin dynamics in the two systems may provide insights for understanding the magnetism in the FeSCs.

In this work, we use X-ray absorption spectroscopy (XAS) and RIXS to study magnetically ordered NaFe$_{1-x}$Cu$_x$As with $x=0$ and $0.47$ [Figs.~\ref{Fig_schematic}(a) and (b)]. Our XAS measurements reveal that Fe atoms maintain a $+2$ valence in both compounds, despite Cu being in a $+1$ state in the latter. Underdamped dispersive spin waves were uncovered in both compounds, consistent with their magnetically ordered ground states. 
Measurements along the $(q,q)$ direction is unaffected by twinning and allows the intrinsic spin wave dispersion to be extracted through RIXS measurements. Compared to spin waves in NaFeAs, the spin waves in NaFe$_{0.53}$Cu$_{0.47}$As are softer in energy but more intense in intensity, indicating a larger total (ordered and fluctuating) moment and stronger electronic correlations associated with its Fe$^{2+}$ state. Our results suggest the magnetism in NaFe$_{0.5}$Cu$_{0.5}$As arises from localized spins, and are consistent with the FeSCs being close to a $n=6$ correlated insulating state.   

\begin{figure*}
	\includegraphics[scale=0.4]{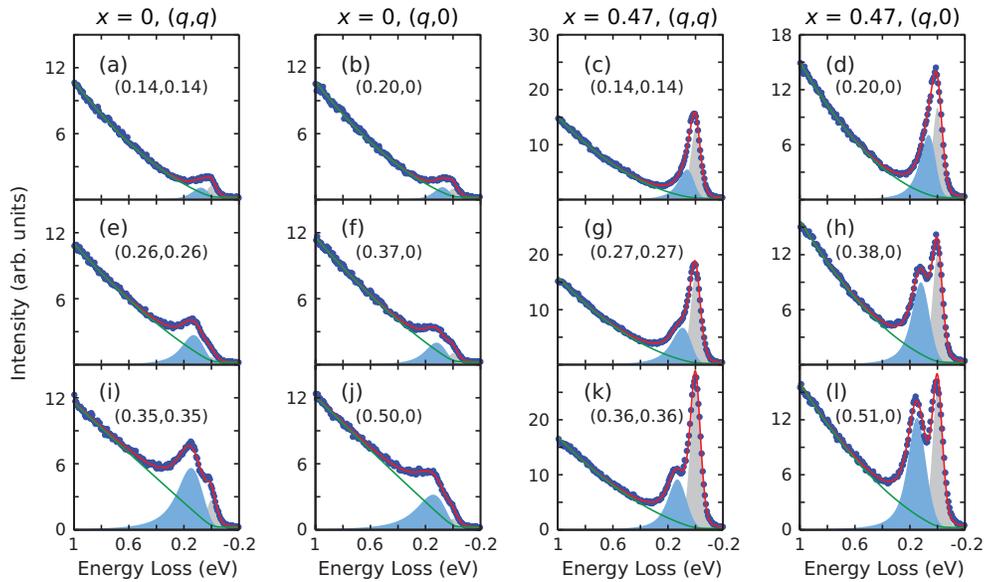} \protect\caption{(Color online) RIXS spectra for NaFe$_{1-x}$Cu$_x$As with $x=0$ and $x=0.47$. Each column of plots corresponds to measurements done for either $x=0$ or $x=0.47$, and for \ys{${\bf Q}_{\rm ab}$} along either $(q,0)$ or $(q,q)$. \ys{${\bf Q}_{\rm ab}$} is labeled inside each panel. \ys{In each panel the measured data points (blue symbols), total fit intensity $I_{\rm total}$ (red line), $I_{\rm bkg}+I_{\rm fluo}$ (green line), $I_{\rm spin}$ (light blue shaded area), and the elastic peak (light gray shaded area) are presented.}
	}
	\label{full_spectra}
\end{figure*}

\section{Methods}
\subsection{Experimental Details} 
Single crystals of NaFe$_{1-x}$Cu$_x$As ($x=0$ and $0.47$) were grown using the self-flux method \cite{YSong2016,AWang2013}. We probe the spin dynamics in the $x=0.47$ sample to approximate the ideal NaFe$_{0.5}$Cu$_{0.5}$As system. RIXS measurements were carried out using the ADRESS beamline at the Swiss Light Source, Paul Scherrer Institute \cite{VNStrocov2010,GGhiringhelli2006}. The instrument energy resolution was calibrated using carbon tape and can be well-described by a pseudo-Voigt form, with a width (FWHM) of $\approx86$~meV. Since NaFeAs and its doped variants are highly air-sensitive, \ys{all sample preparations and transfers were carried out under either Ar or N$_2$ flow}. \ys{Atomically flat pristine} surfaces were then {post-cleaved} inside the chamber under ultra-high vacuum. Momentum transfers are labeled in reciprocal lattice units, using the orthorhombic unit cell [Figs. 1(a) and (b)] with $a\approx b\approx5.59$~{\AA}, $c\approx6.97$~{\AA} for $x=0$ and $a\approx b\approx5.70$~{\AA}, $c\approx6.88$~{\AA} for $x=0.47$. Samples were aligned in either $(H,H,L)$ or $(H,0,L)$ scattering planes. We used a fixed scattering angle of 130$^\circ$, with grazing incident beam predominantly in the $ab$-plane of the sample and linearly polarized in the scattering plane ($\pi$ polarized). The fixed scattering angle results in a fixed total momentum transfer ${\bf Q}$, while the in-plane (${\bf Q}_{ab}$) and out-of-plane (${\bf Q}_{c}$) components can be varied by rotating the sample with respect to the incident beam [Fig. 1(f)]. The accessible range of ${\bf Q}_{ab}$ is limited by kinematic constraints [Fig. 1(e)], and the layered nature of NaFe$_{1-x}$Cu$_x$As samples suggest their spin dynamics have minimal dependence on ${\bf Q}_{c}$ \ys{(for example a dispersion of only a few meV along ${\bf Q}_c$ is seen for NaFeAs \cite{YSong2013b})}. RIXS spectra were collected at the Fe $L_3$-edge ($E_{\rm i}=707.9$~eV). All measurements were carried out at $T=17$~K, well inside the magnetically ordered state of both NaFeAs \ys{($T_{\rm N}\approx45$~K \cite{DParker2009,SLi2009})} and NaFe$_{0.53}$Cu$_{0.47}$As \ys{($T_{\rm N}\approx200$~K \cite{YSong2016})}. The detector used has a pixel density of $\approx30.8$~meV/pixel (with a subpixeling of 4 we have $\approx7.7$~meV/subpixel), providing a significantly finer sampling of the RIXS spectra compared to previous work on NaFe$_{1-x}$Co$_x$As \cite{JPelliciari2016} with $\approx50$~meV/pixel. 

Structural twinning common to the FeSCs with low-temperature orthorhombic lattice structures is expected in both measured samples. In the magnetically-ordered state of NaFeAs, there are magnetic domains that have antiferromagnetic spin stripes running along the $(q,0)$ or the $(0,q)$ direction, corresponding to domains that order magnetically at the $(1,0)$ and $(0,1)$ positions, respectively. Similarly, in NaFe$_{0.53}$Cu$_{0.47}$As, there are domains with Fe chains that run along the $(q,0)$ or $(0,q)$ direction, also resulting in magnetic order at the $(1,0)$ and $(0,1)$ positions, respectively. While the magnetic ordering vectors associated with the two types of domains are separate in momentum space, spin excitations stemming from $(0,0)$ inevitably overlap (Fig. \ref{Fig_schematic}(e)). Therefore, due to twinning, our data measured along the $(q,0)$ direction contain contributions from both the $(H,0)$ and $(0,K)$ directions of a single-domain sample. We note that because each measurement at a particular momentum position is obtained by binning multiple scans taken at randomly selected sample positions, even if structural domain sizes are larger than our beam spot, it remains reasonable to assume that the two domain types contribute similarly to our data. On the other hand, twinning does not affect measurements along the $(q,q)$ direction (Fig. \ref{Fig_schematic}(e)). 
    
\subsection{Data Analysis}
To avoid  beam damage, each measurement is repeated multiple times by translating the sample so that no single position is irradiated for more than 5 minutes. The repeated scans are combined and then normalized to their integrated intensity in the energy loss range $[0.4,10]$~eV, which is dominated by the fluorescence signal, with intensity proportional to the amount of Fe atoms probed in the measurement. Within the energy loss range $[-1,1]$~eV, we modeled the RIXS spectra ($I_{\rm total}$ as a function of energy loss $E$) with a linear background ($I_{\rm bkg}$), an elastic peak ($c\delta(E)$), a fluorescence signal ($I_{\rm fluo}$) and spin excitations ($I_{\rm spin}$):

\beq
I_{\rm total}(E)=I_{\rm bkg}(E)+c\delta(E)+I_{\rm fluo}(E)+I_{\rm spin}(E).
\eeq

\noindent The background $I_{\rm bkg}$ is modeled as a linear function.
The fluorescence is modeled as a second-order polynomial:

\ys{
\beq
I_{\rm fluo}(E)=(p_1 E+ p_2 E^2)H(E),
\label{fluo_eq}
\eeq
}

\noindent
with $H(E)$ being the Heaviside step function.
For consistency, we used the same model to describe the fluorescence for both $x=0$ and $x=0.47$ samples, even though the ground state of NaFe$_{1-x}$Cu$_x$As with $x\approx0.5$ exhibits an electronic gap $\sim100$~meV \cite{CMatt2016,ACharnuka2017}. We have carried out additional analysis to show that our conclusions are robust when $I_{\rm fluo}$ is modified to:  

\ys{
\beq
I_{\rm fluo}(E)=[p_1 (E-\Delta)+p_2 (E-\Delta)^2]H(E-\Delta),
\label{gap_fluo_eq}
\eeq    
}

\noindent
with $\Delta=0.1$ or 0.2 eV (see the Appendix). 

The spin excitations are described by a general damped harmonic oscillator function \cite{JLamsal}:

\beq
I_{\rm spin}(E)=\frac{A}{\pi}\frac{1}{1-e^{-\beta E}}\frac{2\gamma E E_0}{(E^2-E_0^2)^2+(\gamma E)^2},
\eeq

\noindent
which in the limit of no damping ($\gamma\rightarrow0$) and low temperature ($\beta E\gg1$) is $I_{\rm spin}=A\delta(E-E_0)$. In addition, the position of zero energy loss ($E=0$) is also set as a free parameter in our model. The measured RIXS spectra were then fit to our model of $I_{\rm total}$, after convolving with the instrument energy resolution. We then extracted the values and uncertainties (1 s.d.) of the best fit parameters. It should be noted that fit values of $E_0$ and the zero energy position are correlated, so that the uncertainties obtained from fitting may be underestimated. This is especially the case for small in-plane momentum transfers, for which the spin excitations and the elastic line strongly overlap, causing the fit values of $E_0$ to be overestimated and the associated fit uncertainties of $E_0$ to be underestimated.     

\subsection{$U(1)$ Slave-Spin Mean Field Theory}

To understand the nature of the insulating behavior, we studied the multi-orbital Hubbard model for NaFe$_{0.5}$Cu$_{0.5}$As using the $U(1)$ slave-spin theory~\cite{RYu2012}. Since Cu ions exhibit $+1$ valence and are nonmagnetic~\cite{YSong2016}, the local (ionization) potential difference between the Fe and Cu ions is large and will suppress the hopping integral between Fe and Cu sites. This allows us to treat Cu ions in NaFe$_{0.5}$Cu$_{0.5}$As as vacancies, with Fe atoms ordering into the quasi-1D chains as shown in Fig. \ref{Fig_schematic}(b).
For the hopping parameters between Fe ions, we use those for NaFeAs~\cite{RYu2014}. In the calculation, the
electron density is fixed to $n = 6$ per Fe ion, to be consistent with the observed $+2$ valence of Fe. The details of this method is presented in Ref.~\cite{RYu2012}. To study the metal-to-insulator transition of the model, we calculated the quasiparticle spectral weight, $Z_\alpha$, in each Fe $3d$ orbital $\alpha$. The ground state of the system is a Mott insulator if $Z_\alpha=0$ in all orbitals, and is a metal if $Z_\alpha>0$ for all orbitals. We have also found an orbital-selective Mott phase (OSMP), corresponding to $Z=0$ in the $d_{xy}$ orbital, and $Z_\alpha>0$ for other orbitals. The resulting phase diagram is shown in Fig. \ref{theory}.

\section{Results}

\subsection{X-ray absorption spectroscopy}

\ys{XAS near the Fe $L_3$-edge measured in total fluorescence yield (TFY) are compared for NaFe$_{1-x}$Cu$_x$As ($x=0$ and $x=0.47$) in Fig.~\ref{Fig_XAS}(a)}, after normalizing the spectra so that the maximum and minimum intensities respectively correspond to 1 and 0. Both spectra are dominated by an absorption peak at $E=707.9$~eV, without an additional peak at $E\approx710$~eV \cite{HKim2016}. This suggests that Fe ions are in the Fe$^{2+}$ state for both compounds, with minimal contributions from the Fe$^{3+}$ state. This conclusion is surprising for the $x=0.47$ compound, as Cu ions are in the Cu$^{1+}$ state (see Supplementary Information of Ref.~\cite{YSong2016}) and valence counting suggests Fe ions to be in the Fe$^{3+}$ state, when As ions are assumed to be in the As$^{-3}$ state. Since As-As covalent bonding found in $c$-axis collapsed iron pnictides \cite{VKAnand2012} is unlikely for the NaFeAs structure, this suggests that nominally the As ions may be in an unusual As$^{-2.5}$ state. At present, it is difficult to probe directly the valence state of As. The absorption peak at $E=707.9$~eV is slightly sharper for the $x=0.47$ sample compared to $x=0$, which is a result of its more localized nature.

\subsection{Resonant inelastic X-ray scattering}

\ys{Since Fe ions are in the Fe$^{2+}$ ($3d^6$) state for both NaFeAs and NaFe$_{0.53}$Cu$_{0.47}$As, in the RIXS process the electronic configuration of the Fe ions are excited from $2P^{4}_{3/2}3d^6$ to $2P^{3}_{3/2}3d^7$, and then relax back to $2P^{4}_{3/2}3d^6$, allowing for the creation of spin excitations with $\Delta S=1$.}
Fig.~\ref{Fig_XAS}(b) compares representative RIXS spectra for the $x=0$ \ys{(${\bf Q}_{ab}=(0.50,0)$)} and $0.47$ \ys{(${\bf Q}_{ab}=(0.51,0)$)} samples over a broad energy transfer range. As can be seen, broad fluorescence peaks are seen for both samples. The presence of clear differences indicates a significant modification of the electronic structure upon heavy Cu-substitution in NaFeAs \cite{YSong2016,CMatt2016,ACharnuka2017,SZhang2017}. For energies less than 0.5~eV, an additional inelastic peak can be resolved, which can be attributed to spin excitations. The intensity of this feature is more prominent in the $x=0.47$ sample, with the RIXS spectra normalized to integrated intensity in the energy loss range $[0.4,10]$~eV. This suggests that the spectral weight of spin excitations per Fe atom is substantially larger in the $x=0.47$ compound. The elastic peak is also more prominent in the $x=0.47$ sample, \ys{which is likely caused by stronger Laue monotonic scattering \cite{BEWarren} due to increased disorder.} 

To focus on the spin dynamics, we analyzed our RIXS data within an energy window of $[-1,1]$~eV, with representative scans and the corresponding fits summarized in Fig. \ref{full_spectra}. The green solid lines represent the results of fits to the fluorescence on top of a linear background ($I_{\rm bkg}+I_{\rm fluo}$), and the shaded areas are either damped harmonic oscillators that capture the spin excitations ($I_{\rm spin}$) or elastic peaks. As can be seen, all scans are well-described by our model of $I_{\rm total}$, allowing the intensity of the spin excitations $I_{\rm spin}$ to be separated. Fig. \ref{spin_only_spectra} in the Appendix shows a comparison between data and fit $I_{\rm spin}$ after the fit background, elastic peak, and fluorescence have been subtracted.         

\begin{figure}
	\includegraphics[scale=0.65]{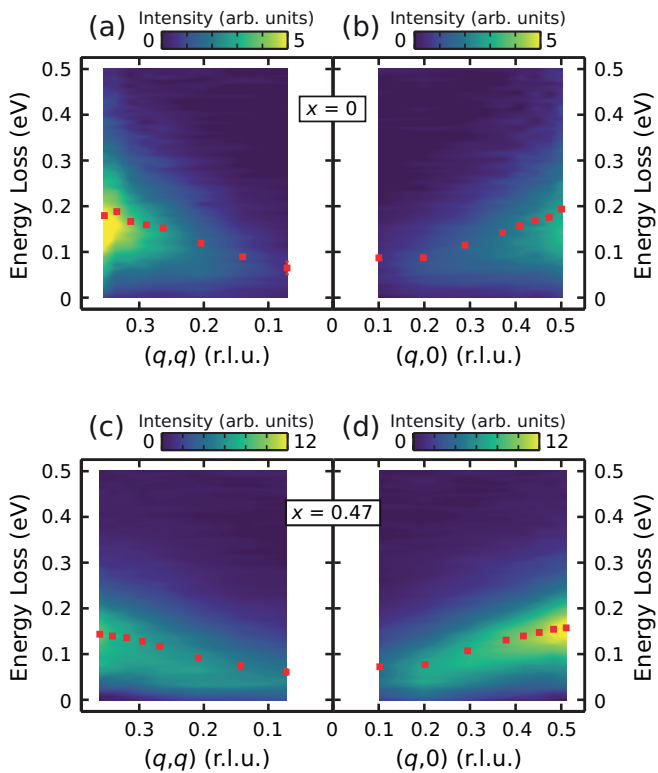} \protect\caption{(Color online) Color-coded and interpolated intensities of spin excitations for NaFeAs along (a) $(q,q)$ and (b) $(q,0)$, and for NaFe$_{0.53}$Cu$_{0.47}$As along (c) $(q,q)$ and (d) $(q,0)$. Red symbols are $E_0$ extracted from our fitting. The intensities of spin excitations are obtained from our data, after subtracting fits to $I_{\rm fluo}$, $I_{\rm bkg}$ and the elastic peak (as done for data in Fig. \ref{spin_only_spectra}).}
	\label{pseudo_color}
\end{figure}
\begin{figure*}
	\includegraphics[scale=0.68]{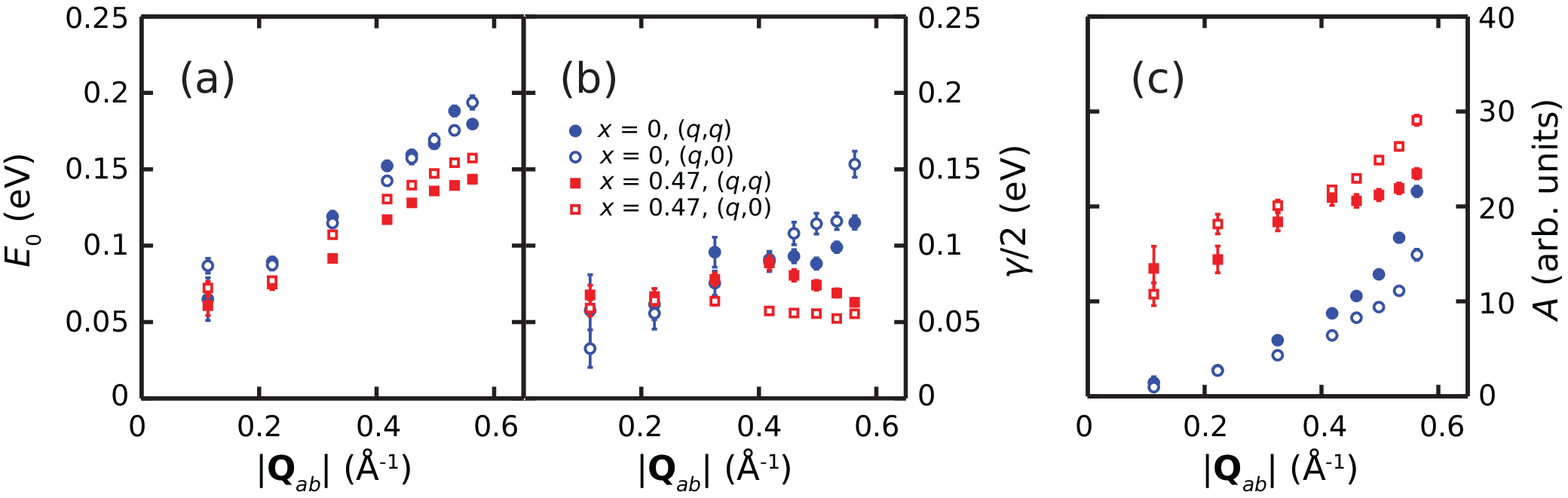} \protect\caption{(Color online) Fit results of (a) $E_0$, (b) $\gamma/2$ and (c) $A$, compared for NaFeAs and NaFe$_{0.53}$Cu$_{0.47}$As.}
	\label{fit_res}
\end{figure*}

Since our data are normalized using the fluorescence peak, it is meaningful to compare quantitatively the intensity of spin excitations as a function of ${\bf Q}_{ab}$ and doping $x$. 
Color-coded and interpolated intensities of spin excitations for the $x=0$ and $0.47$ samples are compared in Fig. \ref{pseudo_color}. Overall, the spin excitations appear more intense in the $x=0.47$ sample (note the different color scales in panels (a)-(b) compared to (c)-(d)).
A quantitative comparison of the spin excitation intensities is presented in Fig. \ref{fit_res}(c), through the fit parameter $A$ ($I_{\rm spin}=A\delta(E-E_0)$ as $\gamma\rightarrow0$). As can be seen, spin excitations are clearly more intense in NaFe$_{0.53}$Cu$_{0.47}$As compared to NaFeAs, with the difference being more prominent near the zone center. The observation of more intense spin excitations in NaFe$_{0.53}$Cu$_{0.47}$As is consistent with its larger ordered magnetic moment \cite{YSong2016}. 
Intensities of spin excitations increase with the increase of momentum transfer, which is a result of RIXS probing the Brillouin zone centered around $(0,0)$. This is distinct from INS measurements, which probe the Brillouin zones near ${\bf Q}_{\rm AF}=(1,0)$ or $(0,1)$, where intensities of spin excitations typically decrease when moving away from ${\bf Q}_{\rm AF}$ \cite{PDai2015}. Moreover, Fig. \ref{pseudo_color} suggests that spin excitations in both NaFeAs and NaFe$_{0.53}$Cu$_{0.47}$As are dispersive, confirmed by energies of $E_0$ extracted from our fits (red symbols). 

To further quantitatively compare the spin excitations in the two studied samples, fit results of $E_0$ and $\gamma/2$ are summarized in Figs. \ref{fit_res}(a)-(b), as a function of the in-plane momentum transfer ${\bf Q}_{ab}$. As can be seen in Fig. \ref{fit_res}(a), spin excitations in NaFeAs are systematically more energetic compared to those in NaFe$_{0.53}$Cu$_{0.47}$As. 
While the dispersions are almost isotropic for NaFeAs along the $(q,q)$ and $(q,0)$ directions, in NaFe$_{0.53}$Cu$_{0.47}$As the dispersion along $(q,0)$ is clearly more energetic compared to $(q,q)$ . $\gamma/2$, which parametrizes damping, is compared in Fig. \ref{fit_res}(b). While similar damping rates are obtained for $|{\bf Q}_{ab}|\lesssim0.3$~${\rm \AA}^{-1}$, spin excitations in the $x=0.47$ sample become clearly less damped compared to those in NaFeAs for $|{\bf Q}_{ab}|\gtrsim0.3$~${\rm \AA}^{-1}$. $E_0$ and $\gamma/2$ in Figs. \ref{fit_res}(a) and (b) are plotted against the same vertical axis, which allows a direct comparison of the two quantities. This is important as spin excitations with a well-defined propagating energy can be assigned only when $\gamma/2<E_0$ (underdamped), since when $\gamma/2>E_0$ the spin excitations become overdamped and are non-propagating \cite{JLamsal}. As can be seen in Figs. \ref{fit_res}(a)-(b), $\gamma/2<E_0$ is satisfied for all measurements in both $x=0$ and $x=0.47$ samples (within fit uncertainties). Our results suggest that although significant damping is present in both samples, the spin excitations remain underdamped, and can be described as spin waves associated with their respective magnetically-ordered ground states.

Given that Fe and Cu atoms order into chains in the ideal NaFe$_{0.5}$Cu$_{0.5}$As structure (Fig. \ref{Fig_schematic}(b)) \cite{YSong2016}, spin excitations in our $x=0.47$ sample should be quasi-1D. The dispersion of spin waves in a 1D antiferromagnetically-ordered spin chain, with two spins in the unit cell along the chain direction (\ys{the $K$-direction in reciprocal space}), is described within classical linear spin wave theory by:
\begin{figure}
	\includegraphics[scale=0.62]{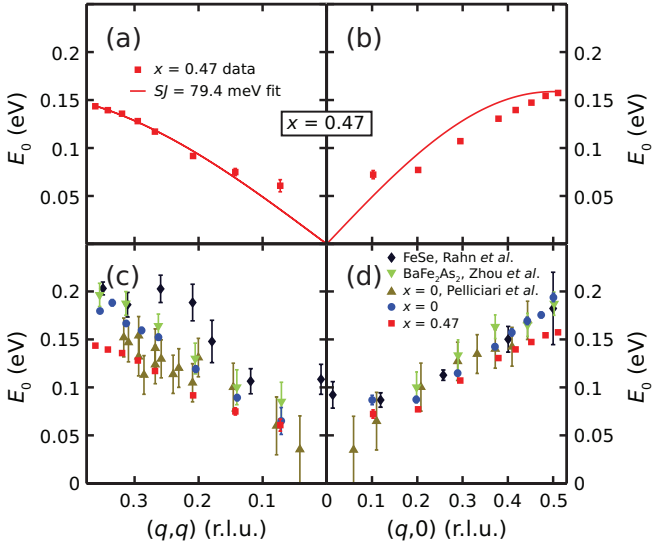} \protect\caption{(Color online) (a) Fit to a classical 1D spin chain model, for the dispersion of spin excitations in NaFe$_{0.53}$Cu$_{0.47}$As along $(q,q)$, resulting in $SJ\approx80$~meV. (b) Comparison of expected dispersion along $(0,K)$ (along the Fe chains) in NaFe$_{0.53}$Cu$_{0.47}$As ($SJ=79.4$~meV) with the measured dispersion along $(q,0)$. Comparisons of the dispersion of spin excitations from our measurements with previous results on BaFe$_2$As$_2$ \cite{KZhou2013}, NaFeAs \cite{JPelliciari2016} and FeSe \cite{CRhan2019}, along (c) $(q,q)$ and (d) $(q,0)$.}
	\label{disp}
\end{figure}
\beq
E({\bf Q})=2SJ\sin(K\pi),
\label{1D_disp}
\eeq 

\noindent with $J$ being the nearest-neighbor exchange coupling and $S$ being the size of the effective spin.
For a quasi-1D system, long-range order is only possible when there are exchange interactions between the chains, which will add $H$ and $L$ dependencies to $E({\bf Q})$. Nonetheless, it is reasonable to assume that $J$ is the dominant interaction and spin waves probed in the energy scale of RIXS are mostly determined by $J$. Given that the $(q,q)$ direction is unaffected by twinning, we fit its dispersion using the 1D model and obtained $SJ=79.4(6)$~meV (Fig. \ref{disp}(a)). This value of $SJ$ is then used to obtain the dispersion along the Fe chain direction (the $(0,K)$ direction, solid line in Fig.~\ref{disp}(b)), in comparison with our measurements along the $(q,0)$ direction (symbols in Fig.~\ref{disp}(b)). As can be seen, the measured dispersion appears significantly lower in energy, which results from twinning. This is because the $(H,0)$ direction is perpendicular to the Fe chains and exhibits spin excitations at considerably lower energies compared to the $(0,K)$ direction, and since our $(q,0)$ measurements contain contributions from both $(H,0)$ and $(0,K)$ directions, the measured energies along $(q,0)$ are lower compared to what's expected for $(0,K)$.
We note that due to strong quantum fluctuations in 1D, while the dispersion of spin excitation is well-described by Eq.~\ref{1D_disp}, the factor 2 is replaced by $\pi$ for $S=1/2$ \cite{JdesColizeaux1962} and \ys{$\approx2.8$} for $S=1$ \cite{SMa1992}. This means that the extracted values of $J$ based on classical linear spin wave theory could be overestimated by $\approx57$\% for $S=1/2$ and $\approx40$\% for $S=1$. In contrast, a much smaller overestimation is expected in two dimensions ($\approx18$\% for $S=1/2$ on a square lattice \cite{RColdea2001}). 

\ys{While spinons dominate the excitation spectrum of an ideal $S=1/2$ antiferromagnetic spin chain, spin waves dominate the magnetically-ordered phase induced by inter-chain interactions \cite{BLake}. The contribution from spinons also becomes less important for $S>1/2$, and if spinons are present in NaFe$_{0.53}$Cu$_{0.47}$As, it would be challenging to distinguish them from the fluorescence.}
\ys{For the points closest to the zone center ${\bf Q}_{ab}=(0,0)$ in Figs.~\ref{disp}(a) and (b), the data appears to deviate from Eq.~\ref{1D_disp}. This may suggest the presence of a sizable zone-center spin gap, or it may result from an overestimate of $E_0$ when its intrinsic value is small (as discussed in Section II. B.). INS measurements which probe spin excitations at the (1,0)/(0,1) zone center would be able to distinguish these two scenarios.}

We compare our RIXS results on NaFe$_{1-x}$Cu$_x$As ($x=0$ and $x=0.47$) with previous results on BaFe$_2$As$_2$ \cite{KZhou2013}, NaFeAs \cite{JPelliciari2016} and FeSe \cite{CRhan2019} in Figs. \ref{disp}(c)-(d). We find our results on NaFeAs to be consistent with previous measurements \cite{JPelliciari2016}, within experimental uncertainties. Along the $(q,q)$ direction, spin excitations in FeSe are the most energetic, followed by those in BaFe$_2$As$_2$, NaFeAs and NaFe$_{0.53}$Cu$_{0.47}$As (Fig. \ref{disp}(c)). Less contrast in the energy of spin excitations is observed 
along the $(q,0)$ direction, with NaFe$_{0.53}$Cu$_{0.47}$As being slightly less energetic than the other compounds (Fig. \ref{disp}(d)).

Starting from NaFeAs and doping Co to span the phase diagram encompassing the superconducting dome,
it was shown that high-energy ($\gtrsim20$~meV) spin excitations exhibit little variation with doping \cite{JPelliciari2016,SCarr2016}. As similar observations were also made for BaFe$_{2-x}$Ni$_2$As$_2$ \cite{HLuo2013}, such a behavior should be general for electron-doped FeSCs. Since Cu dopes electrons near the superconducting dome \cite{SCui2013,YLiu2015}, we expect high-energy spin excitations in superconducting NaFe$_{1-x}$Cu$_x$As to be similar to those in NaFeAs. This means that the softening we observe in NaFe$_{0.53}$Cu$_{0.47}$As likely occurs when doping well beyond the superconducting dome.

\section{Discussion and Conclusion}

As Cu was shown to have $+1$ valence and to be nonmagnetic in NaFe$_{1-x}$Cu$_x$As when $x\approx0.5$ (see Supplementray Information of Ref.~\cite{YSong2016}), valence counting suggests Fe ions to have a $3+$ valence and correspond to a $n=5$ state. The present findings instead indicate Fe ions having a $2+$ valence and remains in a $n=6$ state upon Cu-doping, similar to NaFeAs. This is consistent with the similar local atomic environment found in the two cases (Fig. \ref{Fig_schematic}(c)-(d)), and suggests that As ions have an unusual $-2.5$ valence. Therefore, even though As-As covalent bonding along the $c$-axis as in BaCu$_2$As$_2$ \cite{VKAnand2012} is not possible, a change of As nominal valence in NaFe$_{1-x}$Cu$_x$As occurs upon Cu-substitution (As$^{3-}\rightarrow$As$^{2.5-}$), and is associated with a reduction of the $c$-axis lattice parameter ($c=6.97$~${\rm \AA}\rightarrow c=6.88$~${\rm \AA}$). It appears that the holes contributed through Cu-doping reside on the As sites, rather than the Fe sites. This is analogous to the hole-doped cuprates, in which the doped holes reside on the O sites rather than the Cu sites. This charge transfer physics deserves further theoretical and experimental investigations.

The nearest-neighbor exchange coupling $SJ\approx80$~meV obtained within classical linear spin wave theory for NaFe$_{0.53}$Cu$_{0.47}$As can be compared against effective exchange couplings obtained for the parent compounds of FeSCs. A comparison between the magnetic structures of NaFeAs and 
NaFe$_{0.5}$Cu$_{0.5}$As (Figs. \ref{Fig_schematic}(a)-(b)) shows that the latter can be obtained by replacing half of the Fe ions in NaFeAs by nonmagnetic Cu ions (and an in-plane 90$^{\circ}$-rotation of spins), thus there is a correspondence between $SJ_{1a}$ in NaFeAs and $SJ$ in NaFe$_{0.5}$Cu$_{0.5}$As. $SJ_{1a}$ for archetypal parent compounds of FeSCs are $\approx40$~meV for NaFeAs \cite{CZhang2014}, $\approx59$~meV for BaFe$_2$As$_2$ \cite{LHarriger2011}, $\approx50$~meV for CaFe$_2$As$_2$ \cite{JZhao2009} and $\approx31-39$~meV for SrFe$_2$As$_2$ \cite{REwings2011}. $SJ$ in our $x=0.47$ sample is clearly larger than $SJ_{1a}$ in parent compounds of the FeSCs (the larger bandwidths of spin excitations in the latter include contributions from $SJ_{1b}$ and $SJ_2$), within classical linear spin wave theory. 
The larger value of $SJ$ in NaFe$_{0.53}$Cu$_{0.47}$As could be the result of a larger exchange coupling $J$, a larger $S$ or stronger quantum fluctuations (which leads to a larger overestimation of $SJ$ within classical linear spin wave theory) . To differentiate between these possibilities, it would be desirable to map out magnetic excitations over the Brillouin zone using inelastic neutron scattering, which allows $S$ and $J$ to be independently determined \cite{PDai2015}. Once $S$ is determined, the overestimation of $J$ due to quantum fluctuations can also be corrected for.


Density functional theory (DFT) \cite{YSong2016,SZhang2017} and DFT+dynamical mean field theory (DMFT) calculations \cite{ACharnuka2017} suggest NaFe$_{0.5}$Cu$_{0.5}$As to be metallic in the paramagnetic state, while experimentally semiconducting/insulating transport persists above $T_{\rm N}\approx200$~K. This difference may result from strong electronic correlations not fully captured by DFT and DFT+DMFT calculations \cite{YSong2016,SZhang2017}, which results in NaFe$_{0.5}$Cu$_{0.5}$As being a Mott insulator. Alternatively, it has been suggested that short-range spin correlations persist above $T_{\rm N}$, and NaFe$_{0.5}$Cu$_{0.5}$As is a Slater insulator. Our finding of $SJ\approx80$~meV suggests the dominant exchange interaction energy is substantially larger than $k_{\rm B}T_{\rm N}$, and implies the presence of spin fluctuations above $T_{\rm N}$. Ideally, to differentiate whether NaFe$_{0.5}$Cu$_{0.5}$As is a Mott insulator or a Slater insulator, its transport properties for $k_{\rm B}T\gg J$ should be studied; however, since $k_{\rm B}J$ is comparable to the melting temperature of NaFe$_{0.5}$Cu$_{0.5}$As, this is difficult to achieve experimentally.   

\begin{figure}
	\includegraphics[scale=0.7]{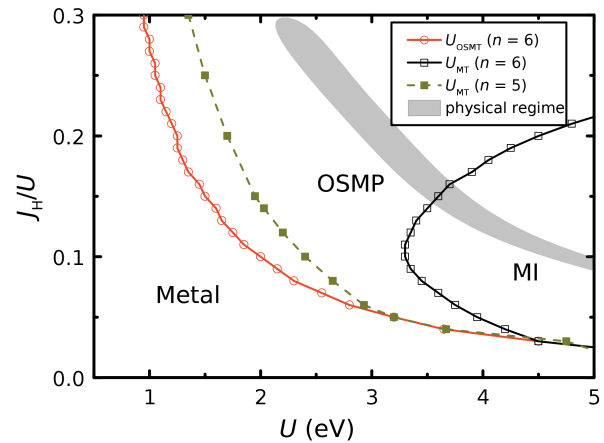} \protect\caption{Theoretical phase diagram of NaFe$_{0.5}$Cu$_{0.5}$As. The shaded gray area is the expected physical regime that the system resides in. Depending on the values of \ys{Hubbard repulsion} $U$ and \ys{Hund's coupling $J_{\rm H}$}, the system may be a metal, an orbitally-selective Mott phase (OSMP), or a Mott insulator (MI). For a fixed $J_{\rm H}/U$, the system may experience an orbitally-selective Mott transition ($U=U_{\rm OSMT}$) and a Mott transition ($U=U_{\rm MT}$), with the increase of $U$. The line of $U_{\rm OSMT}$ for $n=5$ is similar to that of $n=6$ \cite{YSong2016}.}
	\label{theory}
\end{figure}

\begin{figure*}
	\includegraphics[scale=0.4]{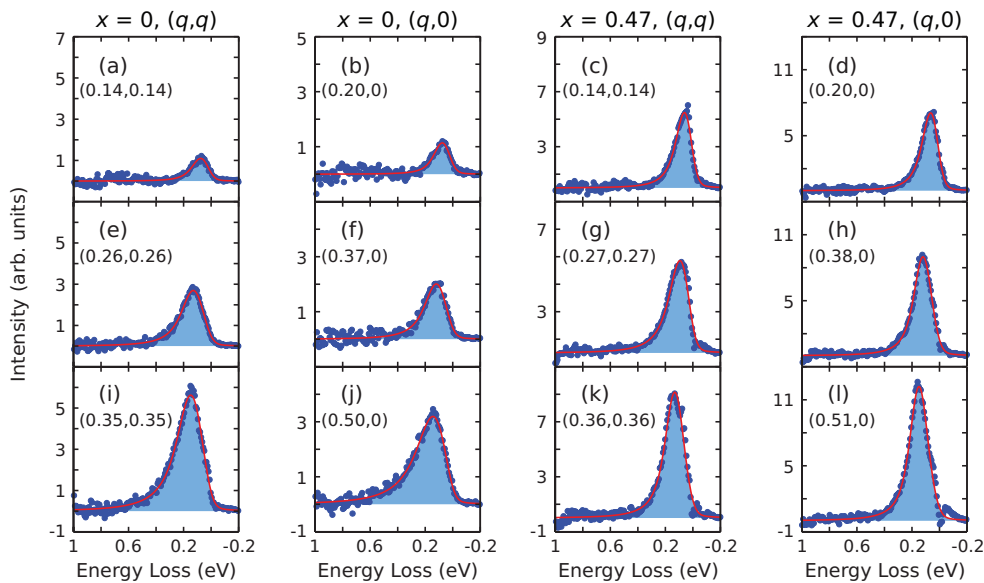} \protect\caption{Similar scans as in Fig. \ref{full_spectra}, but showing only $I_{\rm spin}$ (red lines and light blue shaded areas). Fit values of $I_{\rm bkg}$, $I_{\rm fluo}$ and the elastic peak have been subtracted from the measured data (blue symbols).}
	\label{spin_only_spectra}
\end{figure*}

To shed light on the nature of the insulating state of NaFe$_{0.5}$Cu$_{0.5}$As, we calculated the theoretical phase diagram of NaFe$_{0.5}$Cu$_{0.5}$As using $U(1)$ slave-spin mean field theory (with Fe ions in the $n=6$ state), with results shown in Fig. \ref{theory}. Consistent with previous studies of multiorbital models \cite{LdeMedici2011,LFanfarillo}, we find that for $n=6$ the Mott insulating state survives in a significantly smaller parameter space, compared to half-filling ($n=5$). Using the band renormalization factor obtained from ARPES measurements for NaFeAs \cite{MYi2012}, we estimated the physical regime of $U$ and $J_{\rm H}$ for NaFeAs \cite{YSong2016}. Given the similar local environment and valence of Fe ions in NaFeAs and NaFe$_{0.5}$Cu$_{0.5}$As, this regime of $U$ and $J_{\rm H}$ is then extended to the latter compound. As can be seen, while the physical regime is entirely inside the Mott insulator state of $n=5$, it is only partially inside the Mott insulating state of $n=6$.

Our theoretical results suggest that depending on the precise values of $U$ and $J_{\rm H}$, NaFe$_{0.5}$Cu$_{0.5}$As may either be in a Mott insulating state or a orbital-selective Mott insulating state, with the Slater mechanism relevant for the insulating transport behavior in the latter case. It should be emphasized that in both scenarios, strong electronic correlations are essential for the insulating behavior in NaFe$_{0.5}$Cu$_{0.5}$As. Near the boundary that separates the two cases, we expect a crossover in Mott-type and Slater-type behavior, where distinctions between the two are blurred. 
NaFe$_{0.5}$Cu$_{0.5}$As is thus a rare example of a $n=6$ multiorbital correlated insulator continuously connected to unconventional superconductivity, similar to BaFe$_2$S$_3$ and BaFe$_2$Se$_3$ \cite{HTakahashi2015,JYing2017,SChi2016}. In the latter compounds, the semiconducting/insulating behavior under ambient pressure was attributed to Mott physics \cite{JCaron2012,JRicon2014,MMourigal2015,TYamauchi2015} or a Slater mechanism \cite{SRoh2020}. It is interesting to note the RIXS spectra of  BaFe$_2$Se$_3$ are dominated by $dd$-excitations, which are typically strongly suppressed in the iron pnictides \cite{CMonney2013}, this suggests larger $U$ in BaFe$_2$Se$_3$ compared to the iron pnictides. Similar crossover between Mott-type and Slater-type insulating behavior has been discussed for the iridates Sr$_2$IrO$_4$ \cite{QLi2013,AYamasaki2014,HWatanabe2014} and Nd$_2$Ir$_2$O$_7$ \cite{MNakayama2016}. 

\begin{figure*}
	\includegraphics[scale=0.68]{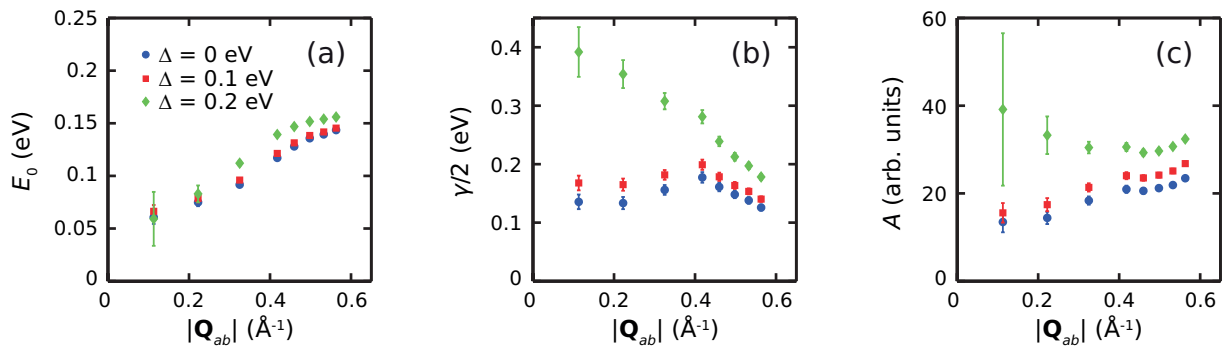} \protect\caption{Comparison of fit results for the fit parameters (a) $E_0$, (b) $\gamma/2$, and (c) $A$, with the introduction of a gap $\Delta$ in $I_{\rm fluo}$ (Eq.~\ref{gap_fluo_eq}). The fits are for data on NaFe$_{0.53}$Cu$_{0.47}$As along $(q,q)$.}
	\label{fluo_gap}
\end{figure*}

As demonstrated using other techniques \cite{IFisher2011}, it should be possible to detwin NaFeAs (or other parent compounds of FeSCs) by applying a small strain in RIXS experiments. For NaFe$_{0.53}$Cu$_{0.47}$As, as twinning is due to Fe-Cu ordering already present at room temperature, it is unlikely strain will have a similar detwinning effect. Nonetheless, it is conceivable that Fe-Cu ordering may disappear at a higher temperature. Applying strain above such a temperature and cooling to room temperature could lead to a detwinned sample. It may be desirable to pursue such a strategy in heavily-doped NaFe$_{1-x}$Cu$_x$As to reveal the system's in-plane anisotropic physical properties using RIXS and other experimental probes. 

In conclusion, we show that Fe ions in NaFe$_{1-x}$Cu$_x$As remains in a $n=6$ state for both $x=0$ and $x=0.47$. RIXS measurements demonstrate the presence of underdamped dispersive spin excitations in both cases. Compared to NaFeAs, spin excitations in NaFe$_{0.53}$Cu$_{0.47}$As are slightly softened in energy but significantly enhanced in intensity. 
In addition, the dispersion of spin excitations is consistent with arising from localized quasi-one-dimensional magnetic moments. 
Our results suggest that NaFe$_{0.5}$Cu$_{0.5}$As likely resides in a crossover regime between Mott-type and Slater-type insulating behavior, and strong electronic correlations are essential for realizing its insulating state.

\section{Acknowledgments} The RIXS work at Rice is supported by the
US Department of Energy (DOE), Basic Energy Sciences
(BES), under Contract No. DE-SC0012311 (P.D.). The materials synthesis efforts at Rice are supported by the Robert A. Welch Foundation, Grant No. C-1839 (P.D.). The work at the University of California, Berkeley and Lawrence Berkeley National Laboratory was supported by the Office of Science, Office of Basic Energy Sciences, Materials Sciences and Engineering Division, of the U.S. Department of Energy (DOE) under Contract No. DE-AC02-05-CH11231 within the Quantum Materials Program (KC2202). J.P. and T.S. acknowledge financial support through the Dysenos AG by Kabelwerke Brugg AG Holding,
Fachhochschule Nordwestschweiz, and the Paul Scherrer Institut.
The synchrotron radiation experiments have been performed at the ADRESS beamline of the
Swiss Light Source at the Paul Scherrer Institut. Part of this research has been funded by the Swiss National Science Foundation through the D-A-CH program (SNSF Research Grant No. 200021L 141325).
The work at PSI is supported by the Swiss National Science Foundation through the NCCR MARVEL and the Sinergia network Mott Physics Beyond the Heisenberg Model (MPBH) (SNSF Research Grants No. CRSII2\_160765/1 and No. CRSII2\_141962).
X.L. acknowledges financial support from the European Community's Seventh Framework Program (FP7/2007-2013) under Grant Agreement No. 290605 (COFUND: PSI-FELLOW).

\section{Appendix}



\subsection{Comparison between data and fit spin excitations}

To allow a direct comparison between our data and fit spin excitations, fits to $I_{\rm fluo}$, $I_{\rm bkg}$ and the elastic peak are subtracted from Fig. \ref{full_spectra} and shown in Fig. \ref{spin_only_spectra}. This allows the fit $I_{\rm spin}$ to be directly compared with experimental data. As can be seen, good fits to the model of $I_{\rm spin}$ have been achieved in all cases.
\subsection{Effect of a gap in the fluorescence on fit results}

Since NaFe$_{1-x}$Cu$_x$As with $x\approx0.5$ is a semiconductor with a lowest electronic gap $\approx0.1$~eV \cite{ACharnuka2017}, we considered modeling its fluorescence using a gapped form (Eq.~\ref{gap_fluo_eq}), which reduces to Eq.~\ref{fluo_eq} when the gap $\Delta$ becomes 0. All of our analysis in the main text were fit using Eq.~\ref{fluo_eq} for $I_{\rm fluo}$. In Fig. \ref{fluo_gap}, we compare fit results of $E_0$, $\gamma/2$ and $A$ with $\Delta=0$, 0.1 and 0.2~eV, for NaFe$_{0.53}$Cu$_{0.47}$As along $(q,q)$. With the increase of $\Delta$, the fit values of $E_0$, $\gamma/2$ and $A$ all increases, with the change being rather small for $\Delta=0.1$~eV. Similar results are also obtained for data along $(q,0)$. 

This analysis suggests that the experimentally obtained charge gap $\approx0.1$~eV is unlikely to significantly affect our fit results. Moreover, our conclusion of enhanced magnetic fluctuations in NaFe$_{0.53}$Cu$_{0.47}$As compared to NaFeAs remains robust with the increase of $\Delta$.

\end{document}